\documentclass{article}
\usepackage{generic}
\usepackage{cite}
\usepackage{amsmath,amssymb,amsfonts}
\usepackage{algorithmic}
\usepackage{graphicx}
\usepackage{textcomp}
\usepackage{lipsum}
\usepackage{hyperref}
\usepackage{subcaption}
\usepackage{float}
\def\BibTeX{{\rm B\kern-.05em{\sc i\kern-.025em b}\kern-.08em
    T\kern-.1667em\lower.7ex\hbox{E}\kern-.125emX}}
    
\author{Maryam Farahmandrad  and  Stefan Goetz}

\begin{document}
\title{Statistical--Spatial Model for Motor Potentials Evoked Through Transcranial Magnetic Stimulation for the Development of Closed-Loop Procedures}

\maketitle

\begin{abstract}
The primary motor cortex appears to be in the center of transcranial magnetic stimulation (TMS). It is one of few locations that provide directly observable responses, and its physiology serves as model or reference for almost all other TMS targets, e.g., through the motor threshold and spatial targeting relative to its position. It furthermore sets the safety limits for the entire brain. Its easily detectable responses have led to closed-loop methods for a range of aspects, e.g., for automated thresholding, amplitude tracking, and targeting. The high variability of brain stimulation methods would substantially benefit from fast unbiased closed-loop methods. However, the development of more potent methods would early on in the design phase require proper models that allowed tuning and testing with sufficient without a high number of experiments, which are time-consuming and expensive or even impossible at the needed scale.  On the one hand, theoretical researchers without access to experiments miss realistic spatial response models of brain stimulation to develop better methods. On the other hand, subjects should potentially not be exposed to early closed-loop-methods without sufficient prior testing as not yet well tuned feed-back as needed for closed-loop operation is known to erratic behavior. 

To bridge this gap, we developed a digital-twin-style population model that generates motor evoked potentials in response to virtual stimuli and includes statistical information on spatial (coil position and orientation) as well as recruitment in the population to represent inter- and intra-individual variability. The model allows users to simulate different subjects and millions of runs for software-in-the loop testing. The model includes all code to stimulate further development.

\end{abstract}


\section{Introduction}
Transcranial magnetic stimulation (TMS) is a noninvasive method to  probe and modulate brain circuits \cite{rossini2015non}. Due to easily accessible and detectable motor responses, the primary motor cortex is one of the most significant targets for brain stimulation. Stimulation of the motor cortex with enough strength can generate motor evoked potentials (MEPs), which can be recorded as a short electric wave through electromyography   at the corresponding peripheral muscle. MEPs have a wide dynamic range from microvolts to millivolts  \cite{hess1987responses, metman1993topographic, matsunaga1998age, paradiso2005representation}. Noninvasive stimulation of the primary motor cortex serves for the diagnosis and localization of motor lesions \cite{miscio1999motor}. Moreover, the primary motor cortex is a preferred model for studying the neurophysiology, biophysics of brain stimulation, and development of novel technology \cite{volz2015makes, goetz2016enhancement, goetz2017development}. 
Most importantly, the motor cortex is a model circuit for most other targets and for safety as well as amplitude individualization through the motor threshold \cite{rossi2009safety}.

MEPs in response to any brain stimulation method depend on a variety of stimulus as well as position parameters and are highly variable between subjects, within subjects from session to session, and from pulse to pulse \cite{ma2024extraction, britton1991variability, kiers1993variability, brouwer1995characteristics, van1996cortical, woodforth1996variability, gur1997response, ellaway1998variability, tobimatsu1998effects, dunnewold1998influence, rosler2008trial,goetz2012model}. One of the most important stimulation parameters with modern focal figure-of-eight coils is accurate targeting, coil placement, and maintenance over time \cite{ngomo2012comparison, cincotta2010optically, gugino2001transcranial, julkunen2009comparison, malcolm2006reliability, weiss2013mapping,goetzkammer2021Oxford}. In addition to coil location \cite{metman1993topographic, ellaway1998variability, mills1992magnetic, schmidt2015nonphysiological}, orientation \cite{richter2013optimal, souza2022tms}, and alignment \cite{koehler2024coil,Koehler2023.11.18.567677}, the distance between the coil and the scalp \cite{richter2013optimal} determine the neuronal response to TMS.

The complexity of MEP formation and the variability responses requires the use of advanced mathematical and statistical methods to precisely and effectively measure parameters such as target hotspot, motor threshold, motor excitability, and neural recruitment input-output curves \cite{meincke2016automated, aonuma2018high, awiszus2003tms, qi2011fast, gotz2011threshold, devanne1997input, carroll2001reliability, pitcher2003age, wang2023three}.  These methods should ideally process MEP responses in real time and adaptively adjust parameters based on previous outcomes, such as determining the next level of stimulation strength for optimal data collection \cite{treutwein1995adaptive, alavi2019optimal}. The development and evaluation of such methods require intensive testing under realistic conditions. It has been shown that widely used methods can be inaccurate and even biased \cite{rossini2015non, awiszus2003tms, gotz2011threshold, westin2014determination}. For closed-loop methods, old experimental data are typically inappropriate as they do not allow sequential adaptive testing of specific parameters but are limited to the available data points. Furthermore, the development of such methods typically requires many participants, test conditions, and repeated trials already during the design phase, typically above a thousand \cite{qi2011fast, gotz2011threshold, awiszus1995comparison, treutwein1999fitting}. Such large experiments are usually not possible in an experimental study, especially in the early development stages of a method.

Closed-loop methods, which automatically tune one or several of the many parameters of a TMS procedure, have moved into the focus of research as they may control the high variability and remove subjectiveness as well as operator influence \cite{LIN202242, lancaster2004evaluation, goetz2019robot}. Furthermore, automated methods can rationalize slow and labor-intensive procedures \cite{mishory2004maximum, van2015tms, HARQUEL2017307, alavi2022formalism}.
Finally, closed-loop operation may enable previously not possible paradigms \cite{GEORGE20231753,singh2023individualized}.

\begin{figure*}[!t]
\begin{center}
\includegraphics[width=0.95\textwidth]{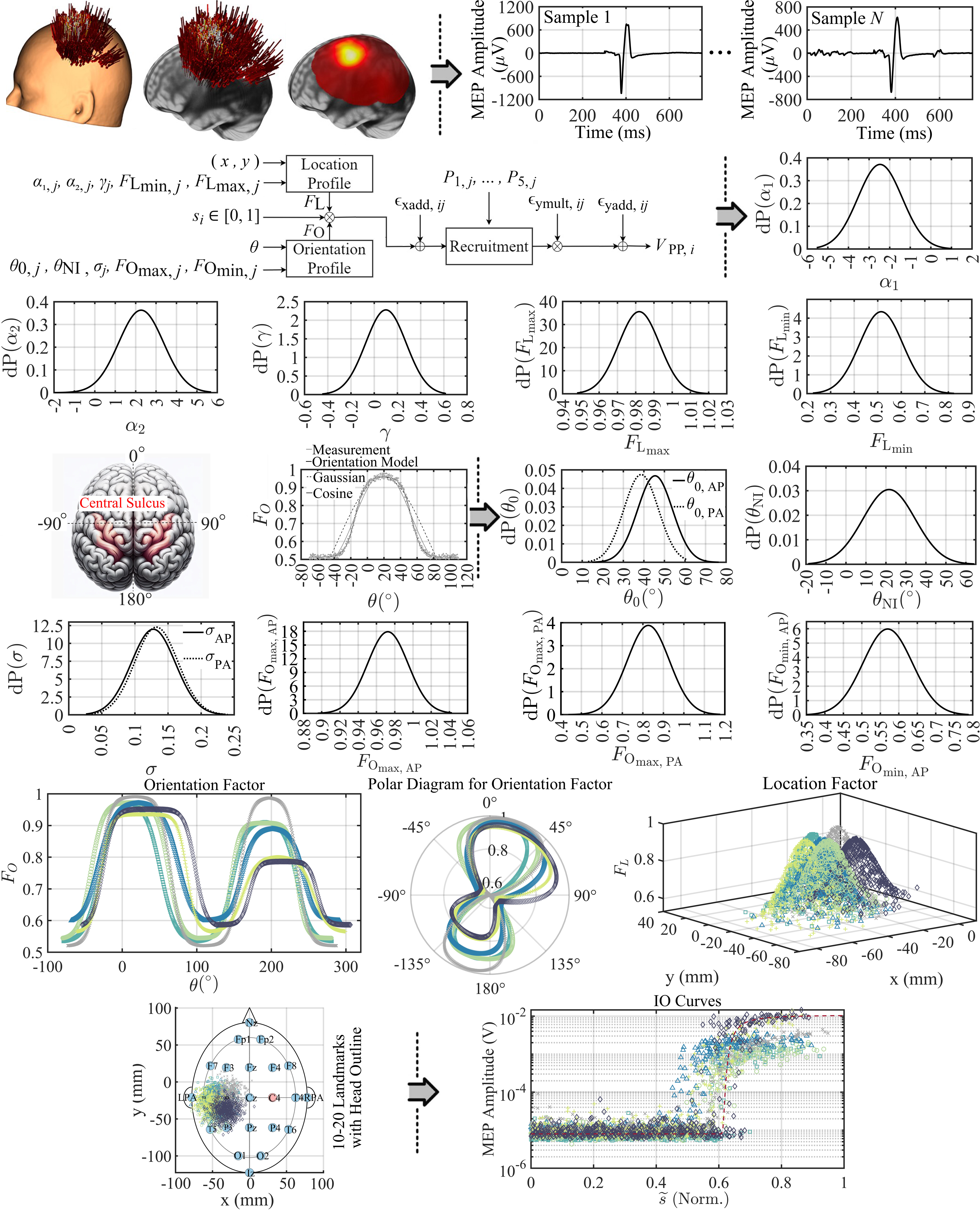}
\end{center}
\caption{Comprehensive visualisation of the spatial--orientation--recruitment model used for closed-loop TMS development.}
\label{FullPage-Figure}
\end{figure*}

This paper developed and provides a  TMS model that generates MEP data in response to stimuli with specific stimulation strength, coil position, and coil orientation in virtual subjects that reflect typical shapes and behavior found in the population. The model incorporates inter-individual variability of spatial location and orientation variability as well as recruitment and intra-individual trial-to-trial variability. As such, it is designed to confront methods during development with a high number of subjects as well as cases and can acquire large numbers of samples per subjects, which are particularly necessary during early methods development when convergence is poor or slow. The model defines virtual subjects from an entire population with their inter-individual variability, each of which further demonstrates intra-individual variability to represent a wide range of physiological responses as known from real-world settings. 

\section{Methodology}
\subsection{Model Data and Model Structure}
We collected data from previous studies, internal databases, and the literature recorded with a range  of devices (Magventure, Magstim, Rogue Research) under neuronavigation on various hand muscles (first dorsal interosseus, FDI; abductor digiti minimi, ADM; abductor pollicis brevis, APB) to cover substantial diversity of conditions and catch a wide range of properties \cite{goetz2019statistical, goetz2016enhancement, goetz2019robot, weiss2013mapping, van2015tms}.

The model is structured into three elements, specifically a spatial model, an orientation model, and an MEP recruitment model.
All model elements are parametric and statistical  so that the combination can describe and represent different subjects. Whereas the spatial and orientation model only contain inter-individuality, the MEP recruitment model further includes intra-individual stimulus-to-stimulus variability. The latter uses a previously developed MEP model \cite{goetz2019statistical}. The parameters of the elements represent an individual subject and can reproduce the same subject later again.
The model furthermore offers a virtual subject generator that includes population statistics for the parameter to randomly generate a virtual subject population where no two subjects are alike.

\subsection{Location Modeling}
Whereas a multivariate normal function did not well represent the spatial map of excitability, particularly due to its symmetry and inability to describe skewness of the profile, we implemented a \textit{multivariate extended skew normal distribution} \cite{tian2016multivariate}. This function represents the spatial model, which provides the effective stimulation strength in the target location for a specific spatial coordinate per
\begin{equation}
\begin{aligned}
F_{\mathrm{L}}(x,y) &= F_{\mathrm{L}_{\mathrm{min}}} + (F_{\mathrm{L}_{\mathrm{max}}} - F_{\mathrm{L}_{\mathrm{min}}}) \\
&\quad \times \left(\frac{2}{2+\gamma}\phi_k(x,y,\boldsymbol{\mu}, \boldsymbol{\Sigma})\Big[1+\gamma\Phi\big(\boldsymbol{\alpha'} (x,y)\big)\Big]\right)\!,
\end{aligned}
\end{equation}
where $F_\mathrm{L}$ represents the location factor that provides the effective stimulus gain at a specific spatial surface coordinate ($x, y$), $F_{\mathrm{L}_{\mathrm{min}}}$ is the baseline location factor, and $F_{\mathrm{L}_{\mathrm{max}}}$ is the maximum location factor.   \( \phi_k(x,y) \) is  the \( k \)-dimensional normal density function with the  mean vector   \( \boldsymbol{\mu} \) and covariance matrix \( \bf{\Sigma} \).  $\Phi(\cdot)$ denotes the cumulative gaussian function, \(\gamma\) the extended parameter, and $\boldsymbol{\alpha} = [ \alpha_1 \: \alpha_2 ]'$  the skewness parameter vector.
$\alpha_1$ and $\alpha_2$ are  the x-direction  skewness and  the y-direction skewness respectively.  

For the population of our experimental study, the parameters $\alpha_1$ ($p = 0.983, \, W = 0.984$) and $\alpha_2$ ($p = 0.964, \, W = 0.982$) show normal distributions. Similarly, $\gamma$ ($p = 0.414, \, W = 0.951$), $F_{\mathrm{L}_{\mathrm{max}}}$ ($p = 0.155, \, W = 0.924$), and $F_{\mathrm{L}_{\mathrm{min}}}$ ($p = 0.645, \, W = 0.964$) also follow normal distributions based on the Shapiro--Wilk test results. These results confirm that these parameters are normally distributed ($p > 0.05$).

Each parameter was represented as a distribution with its respective mean ($\mu$) and variance ($\sigma^2$) as
\begin{equation}
\alpha_1 \sim \mathcal{N}(\mu = -2.50, \sigma^2 = 1.17),
\end{equation}
\begin{equation}
\alpha_2 \sim \mathcal{N}(\mu = 2.23, \sigma^2 = 1.27),
\end{equation}
\begin{equation}
\gamma \sim \mathcal{N}(\mu = 0.100, \sigma^2 = 3.17\times10^{-2}),
\end{equation}
\begin{equation}
F_{\mathrm{L}_{\mathrm{max}}} \sim \mathcal{N}(\mu = 0.982, \sigma^2 = 1.33 \cdot 10^{-4}),
\end{equation}
\begin{equation}
F_{\mathrm{L}_{\mathrm{min}}} \sim \mathcal{N}(\mu = 0.518, \sigma^2 = 8.50\times10^{-3}).
\end{equation}

\subsection{Orientation Modeling}
The orientation model complements the location model with the impact of coil orientation, i.e., rotation around the surface normal, on the recruitment.
The tangential alignment of the coil with the focus point on the local head surface is a question of operator skills rather then a practical degree of freedom in the use of TMS and was studied elsewhere \cite{koehler2024coil,Koehler2023.11.18.567677}.

The orientation model represents the orientation factor that determines the effective stimulus gain as a function of coil orientation. The maximum of the curve typically occurs for coil orientations which generate an induced electric field approximately perpendicular to the central sulcus \cite{souza2022tms}. In this model, the orientation is expressed relative to the nasion--inion line, which serves as a consistent reference based on the 10--20 coordinate system. The angle between the nasion--inion line and the central sulcus varies and depends on the individual anatomy.
Neither previously used cosine \cite{fox2004column} nor Gaussian functions \cite{kallioniemi2015repeatability} provided a good fit to the coil orientation curve.
We identified a generalized logistic function as more appropriate (Figure \ref{FullPage-Figure}), which was previously suggested in the literature \cite{peterchev2013pulse} and describes well the relationship between stimulus strength (input) and the neuronal output. We split the orientation curve into two parts to have individual parameters for the posterior--anterior (PA) and the anterior--posterior (AP) coil orientation and reflect their different shape and excitability \cite{souza2022tms} per
\begin{equation}
F_\mathrm{O}(\theta) = F_{\mathrm{O}_{\mathrm{min}}} + (F_{\mathrm{O}_{\mathrm{max}}} - F_{\mathrm{O}_{\mathrm{min}}}) 
\frac{1}{1 + e^{-\sigma (\theta - \theta_0 - \theta_{NI})}},
\end{equation}
where $F_\mathrm{O}$ represents the orientation factor as a function of coil orientation $\theta$. The terms $F_{\mathrm{O}_{\mathrm{min}}}$ and $F_{\mathrm{O}_{\mathrm{max}}}$ respectively denote the minimum and maximum orientation factors. $\theta_0$ is the center of the logistic response curve,  $\theta_{NI}$ is the angle between the line perpendicular to the central sulcus and the nasion--inion reference line, and \(\sigma\) is the slope of the logistic curve. 

We estimated the parameters separately for AP and PA orientations and tested for normality using the Shapiro--Wilk test. The parameters 
$\theta_{0,\mathrm{AP}}$ ($p = 0.930,\;W = 0.977$), 
$\theta_{0,\mathrm{PA}}$ ($p = 0.546,\;W = 0.953$), 
$\sigma_{\mathrm{AP}}$ ($p = 0.803,\;W = 0.968$) and 
$\sigma_{\mathrm{PA}}$ ($p = 0.451,\;W = 0.948$) show normal distributions. 
Similarly, 
$F_{O_{\max,\mathrm{AP}}}$ ($p = 0.105,\;W = 0.921$), 
$F_{O_{\max,\mathrm{PA}}}$ ($p = 0.139,\;W = 0.928$) and 
$F_{O_{\min,\mathrm{AP}}}$ ($p = 0.347,\;W = 0.949$) also follow normal distributions. 
These results confirm that all tested parameters are normally distributed ($p > 0.05$).

Previous measured coil orientation maxima in the literature do not necessarily agree with each other ($p<0.01$) \cite{reinges2000virtual,hamasaki2012three}. Distributions rarely overlap and the difference between means is larger than, sometimes multiple times, their estimated standard intra-study deviation. Thus, individual studies appear to be under-powered and to under-estimate the real population spread.
We modelled each parameter with individual normal distribution (mean $\mu$ and variance $\sigma^2$) as
\begin{equation}
\theta_{0,\mathrm{AP}} \sim \mathcal{N}\bigl(\mu = 45.5,\;\sigma^{2} = 70.9\bigr),
\end{equation}
\begin{equation}
\theta_{0,\mathrm{PA}} \sim \mathcal{N}\bigl(\mu = 39.1,\;\sigma^{2} = 68.1\bigr),
\end{equation}
\begin{equation}
\theta_{\mathrm{NI}} \sim \mathcal{N}\bigl(\mu = 21.5,\;\sigma^{2} = 169\bigr),
\end{equation}
\begin{equation}
\sigma_{\mathrm{AP}} \sim \mathcal{N}\bigl(\mu = 0.128,\;\sigma^{2} = 1.10\times10^{-3}\bigr),
\end{equation}
\begin{equation}
\sigma_{\mathrm{PA}} \sim \mathcal{N}\bigl(\mu = 0.132,\;\sigma^{2} = 1.00\times10^{-3}\bigr),
\end{equation}
\begin{equation}
F_{O_{\max,\mathrm{AP}}} \sim \mathcal{N}\bigl(\mu = 0.972,\;\sigma^{2} = 5.22\times10^{-4}\bigr),
\end{equation}
\begin{equation}
F_{O_{\max,\mathrm{PA}}} \sim \mathcal{N}\bigl(\mu = 0.827,\;\sigma^{2} = 1.04\times10^{-2}\bigr),
\end{equation}
\begin{equation}
F_{O_{\min,\mathrm{AP}}} \sim \mathcal{N}\bigl(\mu = 0.569,\;\sigma^{2} = 4.40\times10^{-3}\bigr),
\end{equation}
and \(F_{O_{\min,\mathrm{PA}}}\) is obtained by continuity.

\subsection{Integration of Orientation and Location Models into the Recruitment Framework}

 The fundamental recruitment model outputs the peak-to-peak MEP voltage and follows 
\begin{align}
\mathrm{MEP} &= V_\mathrm{PP}(\widetilde{s})\\
&= V_\mathrm{PP}\Big( s \cdot F_L(x,y) \cdot F_O(\theta)\Big),
\end{align}
which combines the spatial $(x,y)$ and orientational ($\theta$) models with the recruitment $V_\mathrm{PP}(\widetilde{s})$ of the effective stimulation strength $\widetilde{s}$ at the target per
\begin{align}
&V_{\mathrm{PP}(\widetilde{s})}\notag&\\ 
&= \epsilon_{\mathrm{yadd},{ij}}(p_{1,j})\notag &\\
&\quad + \exp\left(\ln(10) \cdot \left( \epsilon_{\mathrm{ymult},{ij}} - 7 +  \frac{p_{2,j}}{1 + \frac {p_{3,j}}{\left(\widetilde{s} - p_{5,j}+\epsilon_{\mathrm{xadd},{ij}}\right)^{p_{4,j}}}} \right)\right). 
\end{align}
Stimulation strength $s$ represents the stimulator output.
The parameters $p_{1,j}$ to $p_{5,j}$ are the individual recruitment parameters for Subject $j$. The recruitment is adapted from the literature \cite{goetz2019statistical}.

\section{Combined Model Behavior}
The model allows the generation of virtual subjects whose individual parameters are sampled from the above distributions. The model allows large-scale testing of novel methods at the design stage with very diverse subjects that represent a larger population. The numbers can be several powers of ten larger than a single experimental study may allow. In addition to subject individuality, the responses of a computer-generated virtual subject furthermore demonstrate intra-individual variability with the detailed features known from the literature \cite{kiers1993variability,goetz2019statistical,goetz2022isolating,goetz2014novel,goetz2012model}.

Figure \ref{FullPage-Figure}  visualises of the spatial--orientation--recruitment model. A head model shows the coil-sample positions acquired during data collection, overlaid with a colour-coded motor map where yellow marks the maximum MEP amplitude; representative single-trial MEP waveforms illustrate the raw inputs. 
A block diagram summarises how the location, orientation, and recruitment modules interact to predict the peak-to-peak MEP amplitude. 
The brain sketch illustrates the orientation convention, and a comparison plot contrasts the generalised-logistic orientation model with Gaussian and cosine alternatives. 
The graph at the bottom shows several location and recruitment curves of different virtual subjects with their intra-individual trail-to-trial variability. The parameters were randomly selected based on the identified distributions to represent  a random individual from the population.

The spatial distribution of the location model was visualized within the 10–20 EEG coordinate system, constructed based on a combination of cranial landmarks (nasion, inion, left and right preauricular points). For visualization purposes, the spatial map was projected onto a single brain scalp representation derived from the mean parameters of the virtual subject population. The results demonstrated that the hotspot of cortical excitability was located approximately at the C3 position, corresponding to the primary motor cortex region in the contralateral hemisphere. The spatial maps demonstrate skewness, which was learned from the experimental data and represents the influence of the  cortical gyrification \cite{thielscher2011impact}. The maps furthermore demonstrate peaks, which form the motor hotspot. The orientation leads to two peaks. The one in posterior--anterior direction is higher and not far from 45°, which is well supported by reports from the literature \cite{souza2022tms}. The posterior--anterior peak is smaller and by 180° shifted. The recruitment demonstrates typical sigmoidal behavior with a base noise level with the characteristic skewed extreme-value distribution for low stimulation strength, a rising slope in which the responses rapidly increase on average for stronger stimuli but with high trial-to-trial variability due to an interaction of excitability fluctuations and output variability, and a saturation plateau in which the mostly log-normal variability decreases again. Similar to the shape parameters, also the variability properties are sampled from statistical distributions and vary from subject to subject.

 \section{Conclusion}
This paper introduced an integrated model to support development and test of closed-loop TMS methods on virtual subjects in a software-in-the-loop fashion. The model includes a spatial (cortical coil location and orientation) and recruitment component to generate a matching MEP amplitude for a specific pulse strength administered to a specific location and coil orientation. The model was calibrated to the population represented by study data and represents inter-subject differences in spatial and recruitment behavior. Furthermore, it also includes intra-individual variability and adds  location and orientation into  the recruitment framework. The model successfully simulates virtual subjects with varying anatomical and physiological characteristics, which is crucial for improving model development. The comprehensive    robust framework helps understanding the complex dynamics of brain stimulation.  

\bibliographystyle{IEEEtranTIE}
\bibliography{Manuscript.bib}

\end{document}